\documentclass[a4paper]{jpconf}
\usepackage{graphicx}
\begin{document}
\title{Relevance of nonadiabatic effects in TiOCl}

\author{Diego Mastrogiuseppe and Ariel Dobry}

\address{Facultad de Ciencias Exactas Ingenier{\'\i}a y
Agrimensura, Universidad Nacional de Rosario and Instituto de
F\'{\i}sica Rosario, Bv. 27 de Febrero 210 bis, 2000 Rosario,
Argentina.}

\ead{dobry@ifir-conicet.gov.ar}

\begin{abstract}
We analyze the effect of the phonon dynamics on a recently proposed model for the 
uniform-incommensurate transition seen in TiOX compounds. The study is based on a 
recently developed formalism for nonadiabatic spin-Peierls systems based on bosonization
and a mean field RPA approximation for the interchain coupling. To reproduce the measured 
low temperature spin gap, a spin-phonon coupling quite bigger than the one predicted 
from an adiabatic approach is required. This high value is compatible with the
renormalization of the phonons in the high temperature phase seen in inelastic 
x-ray experiments. Our theory accounts for the temperature of the incommensurate 
transition and the value of the incommensurate wave vector at the transition point.
\end{abstract}

\section{Introduction}
The renewed interest in spin-Peierls (SP) systems arises from the
recently characterized compounds TiOX (X = Cl, Br). TiOCl is a 
quasi-one-dimensional antiferromagnet with strong spin-phonon
coupling. Different than the previously studied organic and inorganic SP
systems, TiOX has an intermediate incommensurate phase between the
dimerized and the uniform ones. Motivated by a phenomenological
Landau-Ginzburg calculation \cite{Ruckamp}, we have recently studied
a model of Heisenberg chains coupled to phonons \cite{MD}. As a
simplified phononic model we had taken arrays of harmonic chains
inserted in an anisotropic triangular lattice. The transversal
elastic coupling arising in this geometry introduces a degeneration
of the phonon mode at the zone boundary (ZB). This degeneration
induces a linear dependency of the phononic dispersion near the ZB
which is different than the usual flat dispersion close to the ZB. 
We had shown that this difference is at the heart of
the fact that TiOX undergoes an instability from the high
temperature uniform phase to an incommensurate phase instead of a
dimerized one.

In spite of the fact that the transition can be accounted by an adiabatic treatment of the phonons, the temperature
scale seems to be strongly renormalized. In fact, by taking the spin-phonon coupling and the bare phonon frequency
obtained at the ZB from ab-initio calculations, neither the phononic dispersion measured at $T=300K$ nor the wave
vector of the incommensuration could be reproduced. By increasing the spin-phonon coupling in order to solve the
previous problem, the critical temperature turned out to be much bigger than the experimental one \cite{MD}. 

In the present work we investigate the consequences of the inclusion
of the dynamics of the phonons. We treat the effective interchain
coupling generated by integrating out the phonons within a  mean
field {\it random phase approximation} (RPA) framework recently developed \cite{DCR}. We show that a
consistent description of the dimerized phase, the renormalization
of the phonons by their interaction with the magnetism, and the
characteristics of the incommensurate-uniform transition are well
accounted within this approach.

\section{\label{model}The model and its low temperature phase}

\begin{figure}[h]
\includegraphics[width=16pc]{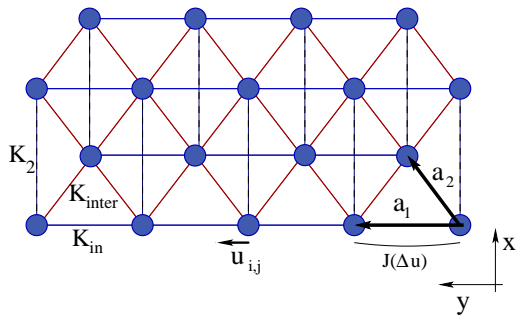}\hspace{1.5pc}
\begin{minipage}[b]{20pc}
\caption{\label{figmodel} Schematic representation of our simplified
model. Only Ti atoms are included over the $xy$ plane. $K_{in}$,
$K_{inter}$ and $K_2$ are the harmonic force constants acting when
two atoms in the same chain, in  neighboring chains
and in next-nearest neighbor chains respectively move from their
equilibrium position. $\textbf{a}_1$ and $\textbf{a}_2$ are the base
vectors. The coordinate of an atom is given by $\textbf{R}_{ij}= i
\textbf{a}_1 + j \textbf{a}_2$.}
\end{minipage}
\end{figure}

We consider the  model proposed in Ref. \cite{MD} for the bilayer of
TiOCl. This simplified model included the spin chains coupled to
phonons and the elastic springs defined on an anisotropic triangular
lattice. Those are the essential ingredients for the incommensurate
transition as previously shown \cite{MD}. The model is
sketched in Fig. \ref{figmodel}. 
Here, we have added a transversal
elastic interaction between ions in second neighbor chains with spring
constant $K_2$. Whereas it does not play any role in the adiabatic
treatment, we will show that it is essential to account for a finite
temperature phase transition in our treatment with nonadiabatic
phonons. The Hamiltonian reads
\begin{eqnarray}
\label{hspinph}
H&=&\sum_{i,j}{\frac{P_{i,j}^2}{2m}} +  \sum_{i,j}\lbrace\frac{K_{in}}{2}(u_{i,j}-u_{i+1,j})^2
+\frac{K_{inter}}{2}\left[( u_{i,j}-u_{i,j+1})^2 +( u_{i,j}-u_{i+1,j-1})^2\right]\nonumber\\
&+&\frac{K_{2}}{2} (u_{i,j}-u_{i-1,j+2})^2 ) \rbrace
+\sum_{i,j} [J + \alpha (u_{i,j}-u_{i+1,j})]\,\textbf{S}_{i,j} \cdot \textbf{S}_{i+1,j},\nonumber
\end{eqnarray}

\noindent where $P_{i,j}$ is the momentum of the atom $i,j$ (the integer indexes identifying an atom in the 
Bravais lattice of Fig.  \ref{figmodel}), $u_{i,j}$ are the displacements from the equilibrium positions along 
the direction of the magnetic chains, $\textbf{S}_{i,j}$  are spin-$\frac12$ operators with exchange constant 
$J=J(\Delta u=0)$ along the y-axis of a non-deformed underlying lattice, 
$\alpha = (\textrm d J(\Delta u)/\textrm d \Delta u)|_{\Delta u = 0}$ is the spin-phonon coupling constant, 
and $K_{in}$, $K_{inter}$ and $K_2$ are the harmonic force constants as shown in Fig. \ref{figmodel}.

We now consider the low temperature dimerized phase. It is customary
to integrate out the phononic coordinates  by going to a path integral
formulation. The generated effective theory contains retarded
interactions between spin-spin dimers along the chain and between different
ones. Its low energy behavior can be studied by bosonization in
similar steps as in Ref. \cite{DCR}. Furthermore, a mean field
treatment of the effective interchain interaction reduces the problem
to a one-chain problem. The mean field order parameter is $\epsilon_0(i)\sim
(-1)^{i} (<\textbf{S}_{i,j}\cdot\textbf{S}_{i+1,j}>-
<\textbf{S}_{i+1,j}\cdot\textbf{S}_{i+2,j}>)$  describing the magnetic
dimerization along the chains. At low temperatures we expect a 
uniformly dimerized phase, thus we take $\epsilon_0$ to be position
independent.
 The resulting one-chain bosonized problem
corresponds to a massive sine-Gordon theory which can be exactly
solved. Solving the mean field equation by minimizing the total
energy with respect to $\epsilon_0$, we obtain for the magnetic
gap
\begin{equation}
\label{deltaJ} \frac{\Delta}{J}= C  F \lambda,
\end{equation}
with
\begin{equation}
\label{F} F(K_{in},K_{inter},K_2) = 1-\frac{1}{\sqrt{1+\frac{K_2}{K_{in}+K_{inter}}}},
\end{equation}
which distinguishes our problem from the square lattice geometry. In Eq.
(\ref{deltaJ}), $\lambda=\frac{\alpha^2 }{J (K_{in}+K_{inter})}$ is  the
adimensional spin-phonon coupling. Note that $(K_{in}+K_{inter})$
appears in $\lambda$ instead of $K_{in}$ alone because it gives the
phononic frequency at $q_y=\pi/b$ for our model.
Furthermore, the constant $C=21.652 \beta^2$ is  model independent. However, it depends on the nonuniversal
constant $\beta$ necessary in the bosonization procedure . We
discuss a procedure to fix $\lambda$ independently of the value of
this constant.

Let us analyze the form  of Eq. (\ref{deltaJ}) for limiting values of
$K_2$ . When $K_2 \rightarrow 0$, then $\Delta\rightarrow 0$, i.e. no gap is
present and the system does not have a dimerized phase because no coherent
dimerization is fixed by the interchain coupling. This is different
than  the results of the adiabatic approach and was the reason to
include $K_2$ in the model. It can be present from the original
microscopic phononic model or it could be generated (at least partially) in
the low energy Hamiltonian as an effective interaction. In the
opposite limit $K_2>> K_{in}+K_{inter}$, we have $\Delta/J = C
\lambda$ which is the result of the adiabatic approach \cite{DCR}.
Let us fix $\Delta/J$ to their experimental values and
look for the spin-phonon coupling in the adiabatic ($\lambda^{ad}$)
and nonadiabatic ($\lambda$) calculations. We have
$\lambda^{ad}/\lambda=F$ depending on the bare phononic
parameters. In the next section we obtain  $F\simeq 0.22$ by
fitting the renormalized phonons to the ones measured in x-ray
experiments at $T=300K$. Moreover $\lambda^{ad}=0.58$ has been
obtained by ab-initio electronic structure
calculations \cite{Valenti}. We obtain $\lambda\simeq 2.6$, thus the
nonadiabatic approach requires a bigger spin-phonon coupling to
produce the same spin gap. We will check this value in the next
section.

Finally, the critical temperature can be obtained as $T_c/J=0.25342 F \lambda$ where $\beta$
has been obtained from the procedure of Ref. \cite{OC}.
We obtain $T_c=96K$ which has to be compared with $T_{c1}=66K$, the
dimerized-incommensurate transition temperature. However, our
$T_c$ corresponds to the stability limit of the dimerized
phase toward an uniform one, and should be taken as an upper limit.
In fact, as we have recently shown for the XY model \cite{MTGD},
our model will undertake a first order phase transition to an
incommensurate phase at a lower temperature than $T_c$. Note that our
value for $T_c$ coincides with the experimental value of $T_{c2}$,
which corresponds to the uniform-incommensurate transition. 

\section{\label{softph}Softening of an incommensurate phonon in the high temperature phase}
\begin{figure}[h]
\includegraphics[width=19pc]{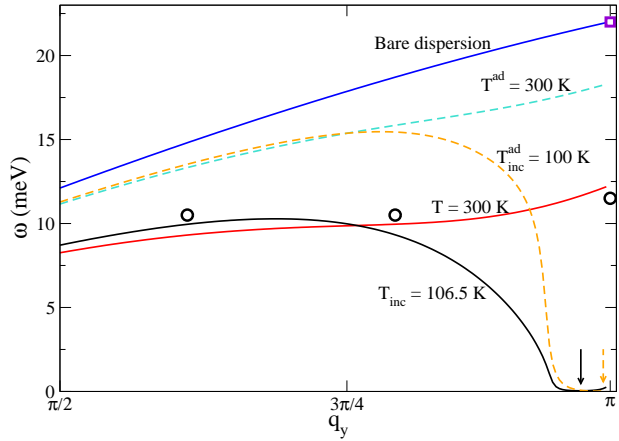}\hspace{1.5pc}
\begin{minipage}[b]{17pc}
\caption{\label{wvsq} Temperature evolution of the dressed phonon frequencies along the ($0,q_y$) path, 
at $T\rightarrow\infty$, $T=300K$ and $T_{inc}$. 
The continuous lines correspond to the nonadiabatic approach while the dashed ones 
correspond to the adiabatic-CF approach. The arrows signal the incommensurate wave vector at the transition point. 
The circles show the experimental frequencies at $300K$ \cite{Abel}.
Note that the bare frequencies are the same in both cases but the spin-phonon interaction is different as 
discussed in the text.}
\end{minipage}
\end{figure}

In the high temperature uniform phase, the order parameter
$\epsilon_0(i)$ vanishes. This is essential to extend the previous
approach to the dynamical correlation functions. Indeed, to compare
with x-ray scattering data, the dynamical structure factor
$S(\textbf{q},\omega)$ has to be determined. It is proportional to
the phononic retarded Green function  $\mathcal{D}^{ret}$. Following
the canonical work of Cross and Fisher (CF) \cite{CF}, we have recently
obtained $\mathcal{D}^{ret}$ by RPA {\it on the spin phonon coupling} \cite{MD}.
This corresponds to fluctuations over the adiabatic approximation for
the phonons. The softening of an incommensurate phonon was obtained,
signaling the uniform-incommensurate transition. In order to go
beyond this approach to discuss the nonadiabatic effects, we use the
result of Ref. \cite{DCR} where the RPA is taken {\it on the
effective interchain coupling} generated by the integration of the
phonons. This follows the same line of reasoning of the previous
section. 
In Fig. \ref{wvsq} we show the evolution with temperature of the dressed phonon frequencies obtained from the 
position of the peaks of $S({\bf q},\omega)$.
The high temperature limit corresponds to the bare phononic dispersion. The frequency at
the ZB, $\sqrt{4(K_{in}+K_{in})/m}$, has been fixed to the one obtained from ab-initio calculations \cite{Valenti2} 
for the degenerate  $A_g$, $B_u$ modes at $175 cm^{-1} (\sim 22meV)$. The relation between $K_{in}$ and $K_{inter}$ and 
the value of 
$K_2$ have been determined by fitting the peaks to the ones measured in inelastic x-ray experiments \cite{Abel} at 
$T=300K$ and the temperature of the incommensurate transition. Our best fit is shown by continuous lines in the 
figure. Once the parameters have been fixed, we are able to calculate $F\simeq0.22$, the value used in
the previous section. We also obtain $q_y \simeq 3.05$ for the incommensurate wave vector at the 
transition temperature which is signaled by the continuous arrow in the figure. This value reproduces very well 
the one measured by elastic x-ray scattering \cite{Abel} for the incommensuration 
in the direction of the magnetic chains.
Note that Fig. \ref{wvsq} has been obtained using the $\lambda=2.6$ obtained from the low temperature phase. 
If, instead of this value, we use $\lambda^{ad}=0.58$ extracted from the adiabatic approach, not a good fitting to 
the measured phonons at $300K$ neither the value of the incommensuration can be obtained. 
The  dashed lines in Fig. \ref{wvsq} show the dressed frequencies obtained from the RPA 'a la' CF, i.e. the adiabatic 
approach, with the previously fitted phononic parameters and $\lambda^{ad}=0.58$.
We see that not a good fitting is obtained.

In summary, we have analyzed the effects of the inclusion of nonadiabatic phonons on a simple model for TiOCl. 
We have found that the spin-phonon coupling is strongly renormalized in relation to the one which arises
from an adiabatic treatment of the phonons. We used the bare phonon frequency obtained by ab-initio calculations 
and fitted the elastic spring constants to obtain the experimental 
phononic frequencies measured at $300K$ by inelastic x-ray scattering. In contrast to the adiabatic treatment,
we obtain a good prediction for the uniform-incommensurate transition temperature 
and for the incommensurate wave vector. 

\ack{This work was supported by ANPCyT (PICT 1647), Argentina.}

\end{document}